\documentclass{llncs}
\usepackage{makeidx}  % allows for indexgeneration
\usepackage{cite} % allows use bibtex
\usepackage{hyperref}
\usepackage[utf8]{inputenc} % input encoding
\usepackage[misc]{ifsym}
\usepackage{tikz}
\usetikzlibrary{arrows,shapes,positioning,shadows,trees,calc,decorations.markings}
\begin{document}
\frontmatter          % for the preliminaries

\mainmatter              % start of the contributions
\title{Ischemic Stroke Lesion Segmentation Using Adversarial Learning} % FIX insert title
\titlerunning{Les_Seg}  % FIX abbreviated title (for running head)
%\author{Ivar Ekeland\inst{1} \and Roger Temam\inst{2} Jeffrey Dean \and David Grove \and Craig Chambers \and Kim~B.~Bruce \and Elsa Bertino}
\author{Mobarakol Islam\inst{1,2} \and N Rajiv Vaidyanathan\inst{2,3} \and V Jeya Maria Jose\inst{2,4} \and
Hongliang Ren \inst{2*}}
\authorrunning{Mobarakol Islam et al.} % FIX abbreviated author list (for running head)
\tocauthor{Example Author} % FIX TOC authors
%
%\institute{Example Institution \email{example@example.org}} % FIX institution
\institute{NUS Graduate School for Integrative Sciences and Engineering (NGS), National University of Singapore, Singapore
\and
Dept. of Biomedical Engineering, National University of Singapore, Singapore\\
\and
Dept. of Mechanical Engineering, NIT, Tiruchirappalli, India\\
\and
Dept. of Instrumentation and Control Engineering, NIT, Tiruchirappalli, India\\
$*$ Corresponding Author: ren@nus.edu.sg\\
\email{mobarakol@u.nus.edu, rajiv.vaidyanathan4@gmail.com, jeyamariajose7@gmail.com, ren@nus.edu.sg }}

%\thanks{I am the corresponding author of the abstract}

\maketitle              % typeset the title of the contribution

\begin{abstract}
Ischemic stroke occurs through a blockage of clogged blood vessels supplying blood to the brain. Segmentation of the stroke lesion is vital to improve diagnosis, outcome assessment and treatment planning. In this work, we propose a segmentation model with adversarial learning for ischemic lesion segmentation. We adopt U-Net with skip connection and dropout as segmentation baseline network and a fully connected network (FCN)  as discriminator network. Discriminator network consists of 5 convolution layers followed by leaky-ReLU and an upsampling layer to rescale the output to the size of the input map. Training a segmentation network along with an adversarial network can detect and correct higher order inconsistencies between the segmentation maps produced by ground-truth and the Segmentor. We exploit three modalities (CT, DPWI, CBF) of acute computed tomography (CT) perfusion data provided in ISLES 2018 (Ischemic Stroke Lesion Segmentation) for ischemic lesion segmentation. Our model has achieved dice accuracy of 42.10\% with the cross-validation of training and 39\% with the testing data. 

\end{abstract}
\section{Introduction}

A stroke occurs due to an interruption in blood flow to the brain. The most common form of stroke is ischemic stroke \cite{world2014cause} which occurs due to a reduction of blood flow, a condition known as ischemia when the brain arteries become narrow or get blocked. It is a medical emergency and undergoes in different disease stages ($acute:0-24h$, $sub-acute:24h-2w$ and $chronic:$ $>$ $2w$) with time \cite{gonzalez2011acute}. At present, Ischemic stroke is assessed by manually delineating the lesion from computed tomography (CT) or magnetic resonance imaging (MRI) including other factors like age, blood pressure, speech, and headache. However, manual delineation is a time-consuming and tedious task which prone to human error and inter-rater variability. Given the severity of the stroke, it is necessary to detect it as quickly as possible. With the recent advancement in the field of deep learning, convolutional neural networks can perform real-time predictions way faster than humans which have improved the process of detection of diseases.

There are few works on developing machine learning and deep learning approaches to detect, localize and segment the stroke lesion. Core and penumbra regions have been characterized using machine learning techniques in sub-acute and acute stage \cite{maier2017isles}. ISLES 2016 and 2017 \cite{winzeck2018isles} presents various deep learning models for stroke lesion segmentation from multi-modal MRI data. A comparison study among different machine learning and convolutional neural network (CNN) has done for stroke segmentation \cite{maier2015classifiers}. Islam et al. \cite{islam2018class} exploits class-balanced PixelNet \cite{bansal2017pixelnet} to segment multi-modal neurological images (MRI) including ischemic stroke in an efficient way. However, these models have not obtained excellent accuracy by considering the visibility of the ischemic stroke in MRI or CT.

In this paper, we propose an architecture for the ischemic stroke lesion segmentation using adversarial learning. The concept of adversarial learning was first introduced in generative adversarial networks ( GANs )\cite{goodfellow2014generative}. It was initially used widely for image synthesis tasks. From then on, the concept of adversarial networks was exploited for usage in a lot of tasks, especially in segmentation. Luc et al. \cite{luc2016semantic} introduce adversarial learning for the semantic segmentation. Consecutively, semi-supervised adversarial learning has been utilized in segmentation which outperforms previous approach \cite{hung2018adversarial}. In this work, we have proposed a variant of the adversarial network for segmentation problems by inspiring \cite{luc2016semantic, hung2018adversarial}. We have also discussed its performance for the segmentation of ischemic stroke lesion.

\section{Proposed Model}
In this section, we discuss about the in detail architecture of the proposed model. The model consists of two deep networks working simultaneously, a concept adopted from the generative adversarial network\cite{goodfellow2014generative} and adversarial learning for segmentation \cite{luc2016semantic, hung2018adversarial}. We simultaneously train two models:  a generative model G that generates the synthesized model, and a discriminative model D that estimates the probability that a sample came from the original data rather than G. The training procedure for G is to maximize the probability of D making a mistake. In our proposed model, the generator is replaced with a segmentor so that it can take in input CT data and produce segmentation label maps as its output. The discriminator is not changed.
%\begin{figure}[!htbp]
\begin{figure}[!t]

\centering
\includegraphics[width=.8\textwidth]{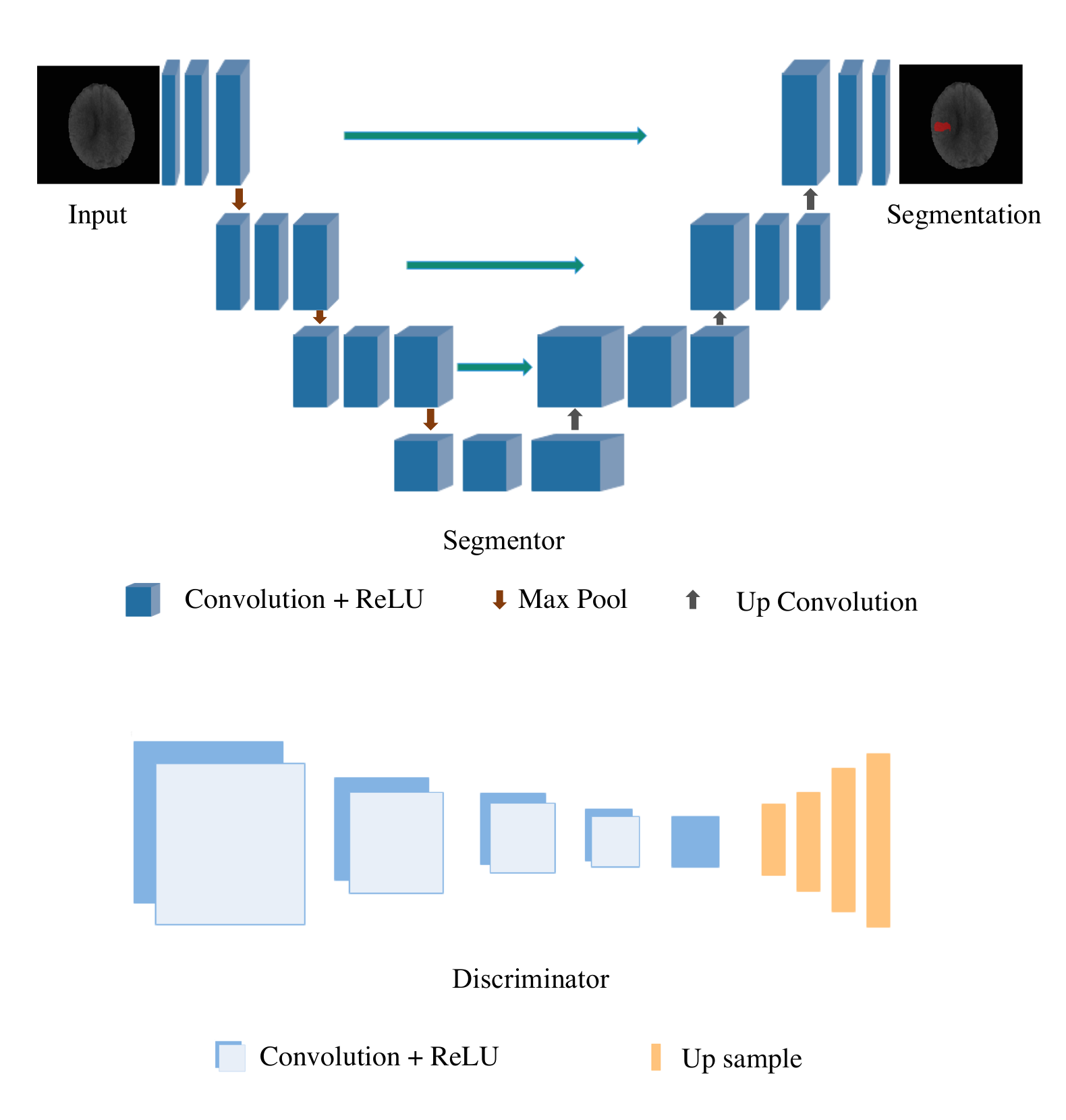}
\caption{UNet for Adversarial Training to segment Ischemic Stroke from CT images.}
\label{fig:Architecture}
\end{figure}
\subsection{Segmentor}
U-Net \cite{drozdzal2016importance, ronneberger2015u} has been used as the segmentor. The network architecture has been illustrated in Fig. \ref{fig:Architecture}. It is of an architecture wherein the convolution layers in the left side contract the input image and the right side expands the feature maps produced by the last convolution layer in the left side. In the left side, the convolutional layers are of 3x3 kernel size with stride 1 followed by a rectified linear unit (ReLU)\cite{nair2010rectified}. A 2x2 max pooling layer is used for downsampling with a stride 2. We double the number of feature channels at the same downsampling stage. In the right side of the network, upsampling layers are illustrated. They follow a 2x2 up-convolution which reduces the feature maps by half. A kernel size of 1x1 has been utilized at the final layer to produce feature maps of desire number classes.
At the final layer, a 1x1 convolution is used to map each 64 component feature vector to the desired number of classes. In total the network has 23 convolutional layers.
\subsection{Discriminator}
Fully Convolutional Network\cite{hung2018adversarial} is used as the Discriminator. It consists of four layers of 2D Convolution followed by ReLu activation. The number of channels increases in each convolution. In the final convolution layer, 512 channels are convolved into 2 confidence maps which consist information about the probability of real or fake data. These confidence maps are then up-sampled to the original size where each upsampling layer has a scale factor of 2.

\section{Experiments}
\subsection{Dataset}

We use ISLES 2018 (Ischemic Stroke Lesion Segmentation) CT dataset to perform all the experiments. ISLES 2018 consists 6 modalities of CT perfusion images. Such as CT, DPWI, CBF, CBV, MTT, and Tmax. Each modality has 3D CT scan of 256x256 axial dimension and a variable slice of 2 to 18 (approximately). There are 94 and 62 cases for training and testing set respectively.  The ground-truth for 94 training cases is given in the training phase. The annotation contains 2 classes where 1 and 0 denote stroke lesion and the healthy tissue respectively. 

\subsection{Training}
We exploit only 3 modalities namely CT,DPWI and CBF. We split the training data into train and valid data in an 80/20 ratio to find the best epoch for segmentation. For testing, another 62 cases are used.
The 3D CT image is first converted into stacks of 2D CT images. These images are given as input to the segmentor which gives probability maps of segmented image as its output. The synthesized segmented probability maps from generator is fed to the discriminator as one of the inputs along with the ground truth. The discriminator is simultaneously trained on the ground truth which is labelled as 1 corresponding to real. 0 corresponds to fake data which is the initial data produced by the segmentor. So, when the probability maps from generator are fed to the discriminator, it gives confidence maps from which the cross-entropy loss is found with respect to two classes. With this adversarial loss along with the segmentor's loss, we seek to train the segmentation network to fool the discriminator by maximizing the probability of the segmentation prediction\cite{hung2018adversarial}. The total loss function is denoted as

\[ \chi = \chi_{seg} + \lambda_{adv}\chi_{adv}  \]

where \(\chi\) is the total loss that is used to backpropagate through both the segmentor and the discriminator. \(\chi_{seg}\) is the loss that is got in the segmentor by comparing the output probability map directly with the ground truth. \(\chi_{adv}\) is the loss that is got in the discriminator which is actually the cross-entropy loss calculated between the two classes. \(\lambda_{adv}\) is the factor introduced to increase or decrease the amount by which adversarial loss will affect the total loss. \(\lambda_{adv}\) for ischemic stroke lesion data set is taken as 0.1. The total loss is backpropagated in both the segmentor and discriminator.

\subsection{Testing}

While testing, the discriminator network is removed and the segmentation labels are got by inputting the test image to the segmentor network. 2D slices have been fed to the model to predict lesion. Finally, all the predicted slices have converted to 3D CT to measure the performances. 
\section{Results}
\begin{figure}[!htbp]
\centering
\includegraphics[width=0.5\textwidth]{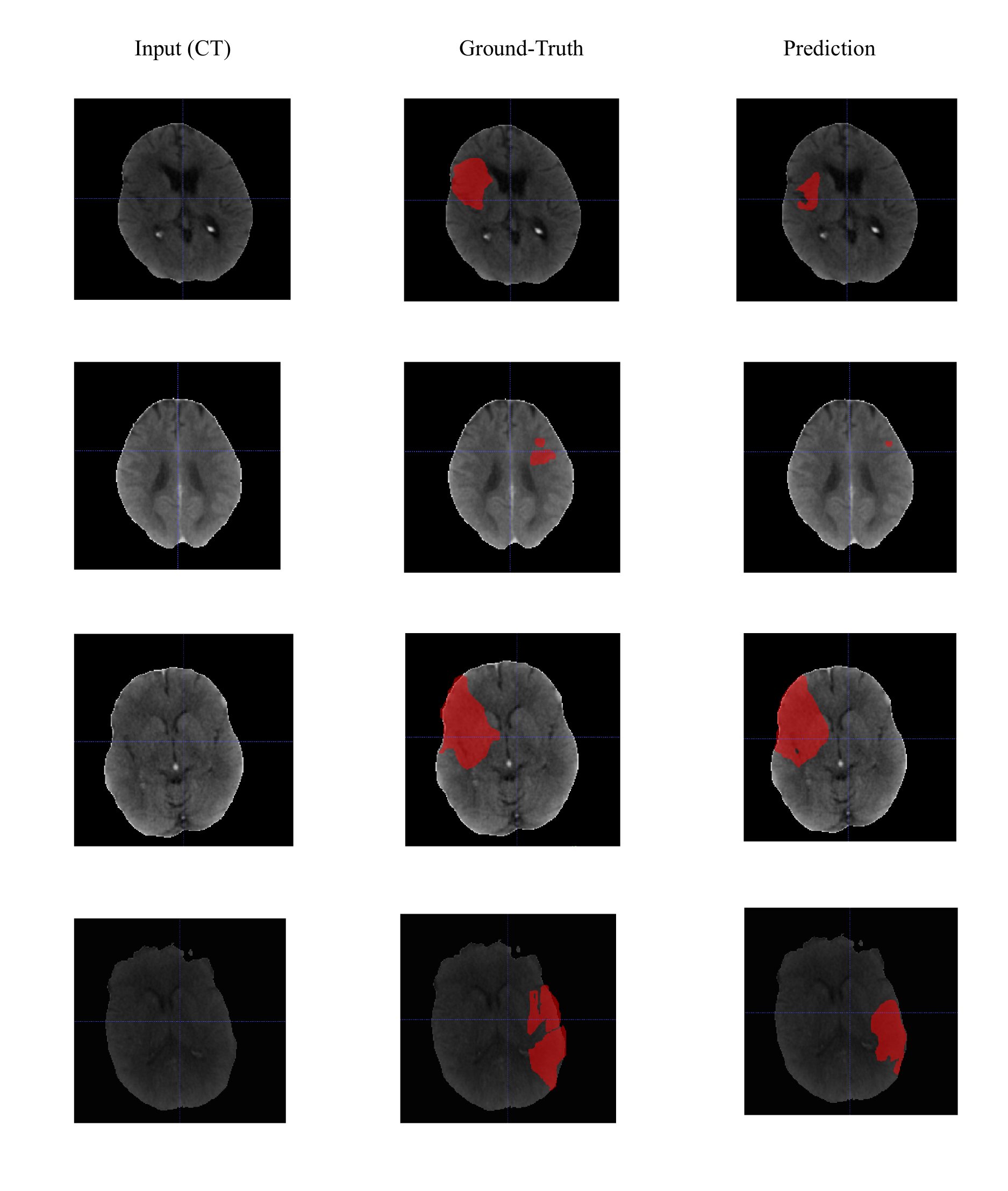}
\caption{Qualitative results of cross validation on training data}
\label{fig:Architecture1}
\end{figure}

\begin{figure}[!htbp]
\centering
\includegraphics[width=0.5\textwidth]{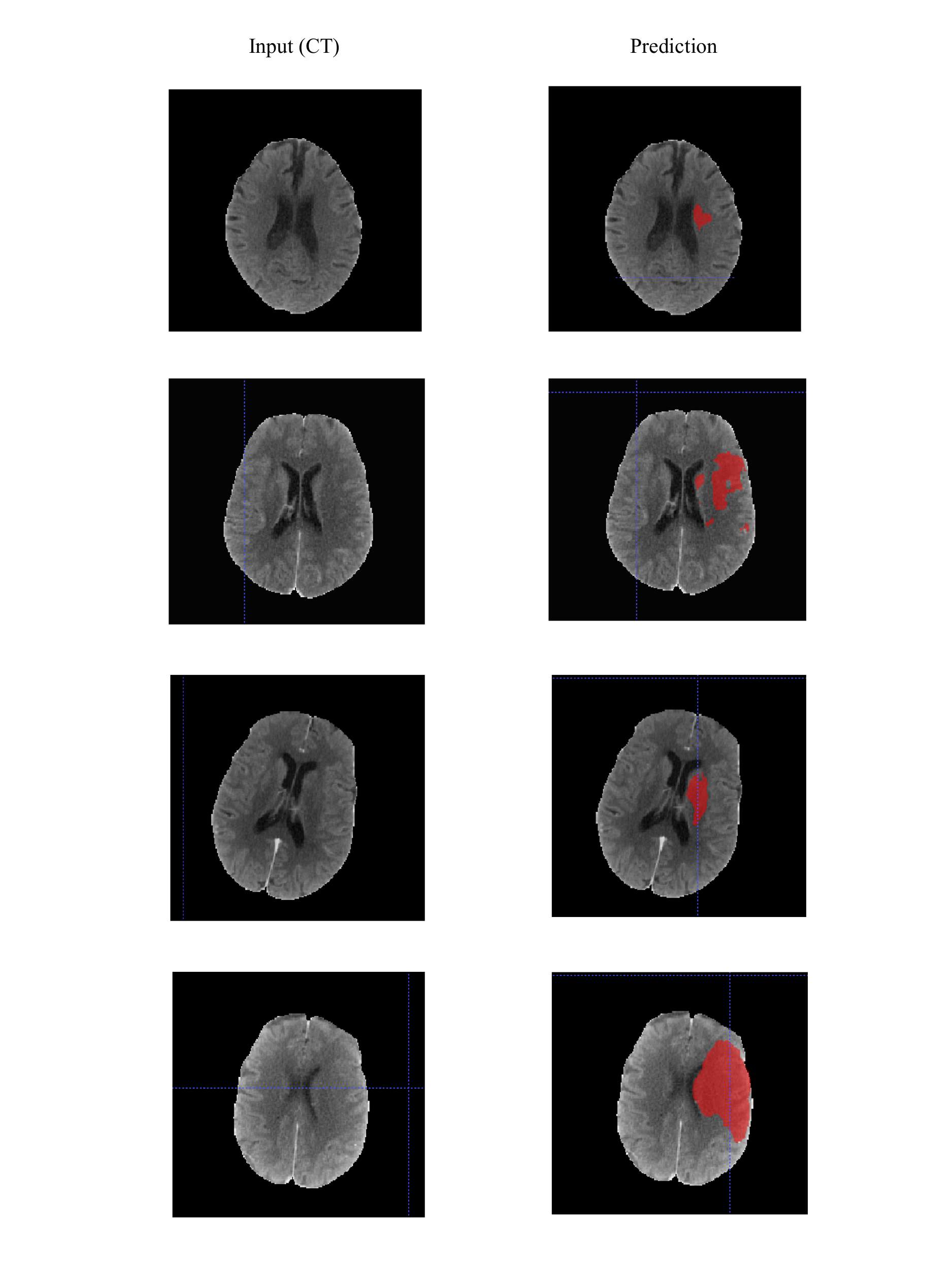}
\caption{Qualitative results of on testing data}
\label{fig:Architecture2}
\end{figure}

The cross-validation dice accuracy of different segmentation architectures has been compared with our proposed model in Table \ref{table:cross_valid}. Table \ref{table:testing} gives details about the performance of our model on the test data set. Fig. \ref{fig:Architecture1} and Fig. \ref{fig:Architecture2} show the predictions of the proposed model for the validation and test data set respectively.

\begin{table}[!h]
\caption{Cross validation performance of different models with training dataset}
\begin{center}
\label{table:cross_valid}
\begin{tabular}{|l|c|c|c|c|c|c|}
\hline
Models & Ours &PixelNet \cite{islam2018class} &U-Net\cite{ronneberger2015u} &Deeplab v2 \cite{chen2018deeplab} &ICNet\cite{zhao2017icnet} &PSPNet\cite{zhao2017pyramid} \\ \hline
Dice & \textbf{0.421} &0.409 & 0.419 & 0.373 & 0.387 & 0.319 \\ \hline
\end{tabular}
\end{center}
\end{table}

\begin{table}[!h]
\caption{Performance of our model with testing dataset}
\begin{center}
\label{table:testing}
\begin{tabular}{|l|l|l|l|l|l|l|}
\hline
  Dice & Hausdorff &Avg Distance &Precision &Recall &AVD\\ \hline
  0.39 & 17741954.64 & 17741938.19 & 0.55 & 0.36 &10.90 \\ \hline
 
\end{tabular}
\end{center}
\end{table}

\section{Discussion}
From the above results, it is evident that even though the network gives a good segmentation prediction, there is still room for improvement. With the same architecture, the network would perform better with a more consistent data set. This data set is inconsistent and has certain defects as some data had just two 2D slices and some images were of poor contrast or shaken. The data set is still less in number as it is difficult to find the best-converged network with such fewer cross-validation data. Even the performance gap in different folds of the cross-validation is very high.

From Table \ref{table:cross_valid}, it can be noted that U-Net, when trained in an adversarial way, gave a better dice accuracy when compared to normal U-Net. Training a segmentation network along with an adversarial network can detect and correct higher order inconsistencies between the segmentation maps produced by ground-truth and the Segmentor. This is due to the extra addition to the loss function that is given due to adversarial learning. Adversarial loss improves the performance of both the segmentor and discriminator. As the training is similar to a min-max game and as the segmentor does not only have its loss but also the extra loss from the discriminator, adversarial training of any segmentation architecture gives a better result when compared to its normal training.

\section{Conclusion}
In this paper, we have proposed an automatic ischemic stroke lesion segmentation using adversarial learning scheme. We have demonstrated than adversarial learning can improve the performance of the segmentation architecture like U-Net \cite{ronneberger2015u}. Our model has achieved better performance comparing to other state of the art models. However, the overall performance of the model is not excellent because of poor visibility of the lesion area in the CT imaging. Moreover, 3D deep learning models cannot be suitable for this dataset where dept slices are inconsistency and very small. As a future work, we can reslice or resample the CT into a higher dimension and try to exploit 3D models directly on 3D CT images.

\section*{Acknowledgement}
This work is supported by the Singapore Academic Research Fund under Grant {R-397-000-227-112}, NUSRI China Jiangsu Provincial Grant BK20150386 and  BE2016077 and NMRC Bedside \& Bench under grant R-397-000-245-511 awarded to Dr. Hongliang Ren.

\bibliography{mybib}{}
\bibliographystyle{plain}

\end{document}